\documentstyle[aps, manuscript]{revtex}
\tightenlines
\draft
\begin{document}
\title{Single field inflationary models with
non-compact Kaluza-Klein theory}
\author{$^1$Diego S. Ledesma\footnote{
E-mail address: dledesma@mdp.edu.ar}
and $^{1,2}$Mauricio Bellini\footnote{
E-mail address: mbellini@mdp.edu.ar}}
\address{$^1$Departamento de F\'{\i}sica, Facultad de Ciencias
Exactas y Naturales
Universidad Nacional de Mar del Plata,
Funes 3350, (7600) Mar del Plata, Buenos Aires, Argentina.\\
and \\
$^2$ Consejo Nacional de Investigaciones Cient\'{\i}ficas y
T\'ecnicas (CONICET)}

\vskip .5cm
\maketitle
\begin{abstract}
We discuss a semiclassical treatment to
inflationary models from Kaluza-Klein theory without
the cylinder condition. We conclude that the evolution of the
early universe could be described by a geodesic trayectory of a
cosmological 5D metric here proposed, so that the effective 4D
FRW background metric should be a hypersurface on
a constant fifth dimension.
\end{abstract}
\vskip .2cm                             
\noindent
Pacs numbers: 04.20.Jb, 11.10.kk, 98.80.Cq \\
\vskip 1cm
\section{Introduction and motivation}

In the last years
cosmological models with extra dimensions has been studied
by many authors\cite{1} with different approaches. One of these
is the Space - Time -Matter theory (STM) developed by P. Wesson and
co-workers\cite{2}, which is one of the versions
of the Kaluza-Klein (KK) theory.
There are three versions of the Kaluza's theory. The first one is known as
compactified KK
theory. In this approach, the Kaluza's cylinder condition
is explained through a physical mechanism of compactification for the
fifth dimension proposed by Klein. In the second version this condition
is explained using projective geometry, in which the fifth dimension
is absorbed into ordinary 4D spacetime provided the (affine) tensors
of general relativity are replaced with projective ones\cite{pdl1}.
In the third version the cylinder condition is not imposed and there
are no assumptions about the topology of the fifth dimension.
This is the usual scenario in non-compact KK theories.

In the STM
theory of gravity the 5D metric is an exact solution of the 5D field
equations in apparent vacuum\cite{3}. The interesting here, is that
matter appears in four dimensions without any dimensional compactification,
but induced by the 5D vacuum conditions. In this framework,
the study of the early universe has great interest.
The equivalence between STM theory and brane-world (BW) theories\cite{bw} has been
studied recently\cite{pdl}. In BW theories the usual matter in 4D is a
consequence of the dependence of 5D metrics on the extra coordinate.
If the 5D bulk metric is independent of the extra dimension, then the
brane is void of matter. Thus in brane theory, matter and geometry are
unified.
In particular, in this letter
we are interested in the study of inflationary models
from the STM formalism. Inflationary cosmology has been studied
from the STM formalism for de Sitter (with a scale factor
that evolves as $ a(t) \sim e^{H_0 t}$)
and power-law inflation (for $a \sim t^p$)\cite{3,4}
using respectively the metrics
\begin{eqnarray}
&& dS^2 = \psi^2 dt^2 - \psi^2 e^{2 \sqrt{\Lambda/(3\psi^2)} t} dR^2
- d\psi^2, \label{1} \\
&& dS^2 = \psi^2 dt^2 - \psi^{2p/(p-1)} t^{2p} dR^2 - \frac{t^2}{(p-1)^2}
d\psi^2. \label{2}
\end{eqnarray}
In metric (\ref{1}), the Hubble parameter is given by the
cosmological constant
and the fifth coordinate: $H_0^2=\Lambda/(3\psi^2)$.
As has been demonstrated, both metrics describe inflationary expansions on a
4D space-time embedded
in a 5D manifold with $\psi$ constant\cite{5}.
These 5D metrics
on comoving spatial coordinates and constant $\psi$ has an interval
given by $dS =\psi dt$\cite{6}.
This should be consistent with
4D particle dynamics, whose corresponding interval or action is
defined by $ds=m dt$. So, if $\psi$ is constant the rest mass $m$ of
a given particle should be constant in this particular frame.
However, we could choose a frame in which $\psi $ varies and hence
the mass
of the 4D particle were variable. For example, as was demonstrated
in \cite{6}, by means of the
5D geodesic equation
\begin{equation}\label{5}
\frac{dU^C}{dS} + \Gamma^C_{AB} U^A U^B =0,
\end{equation}
we can see that in the metric (\ref{2}) we obtain the temporal dependence
of the fifth coordinate: $\psi(t) = (t/t_0)^{(p-1)^2}$ when the spatial
velocities $U^1=U^2=U^3=0$. Here, $\Gamma^C_{AB}$ ($A,B,C=0,1,2,3,4$)
are the 5D Christoffel
symbols
and the velocities are given by $U^A = dx^A/dS$.
From the point of view of a 5D general relativity theory (which we are
working here), it implies that the action is minimized
in this particular frame.

The metrics (\ref{1}) and (\ref{2}) are the 5D extension of a
4D spatially isotropic, homogeneous and flat Friedmann-Robertson-Walker (FRW)
spacetime.
These can be written in a more general manner\cite{5}
\begin{equation}\label{1'}
dS^2 = - e^{\alpha(\psi,t)} dt^2 + e^{\beta(\psi,t)} dR^2 + e^{\gamma(\psi,t)}
d\psi^2,
\end{equation}
where $dR^2 = dx^2+dy^2+dz^2$ and $\psi$ is the fifth coordinate.
The equations for the relevant Einstein
tensor elements are
\begin{eqnarray}
G^0_{ \  0}& =& -e^{-\alpha} \left[\frac{3 \dot\beta^2}{4} +
\frac{3\dot\beta \dot\gamma}{4}\right] + e^{-\gamma}\left[
\frac{3 \stackrel{\star\star}{\beta}}{2}
+\frac{3\stackrel{\star}{\beta}^2}{2}-\frac{3\stackrel{\star}{\gamma}
\stackrel{\star}{\beta}}{4}\right],\\
G^0_{ \  4} & = & e^{-\alpha}\left[\frac{3\stackrel{\star\cdot}{\beta}}{2}
+\frac{3\dot\beta
\stackrel{\star}{\beta}}{4}
-\frac{3\dot\beta \stackrel{\star}{\alpha}}{4} - \frac{3
\stackrel{\star}{\gamma} \dot\gamma}{4}\right],\\
G^i_{ \  i} & = &- e^{-\alpha} \left[\ddot\beta + \frac{3\dot\beta^2}{4}+
\frac{\ddot\gamma}{2} + \frac{\dot\gamma^2}{4} + \frac{\dot\beta\dot\gamma}{2}-
\frac{\dot\alpha\dot\beta}{2} - \frac{\dot\alpha\dot\gamma}{4}\right] \nonumber \\
& + & e^{-\gamma}\left[\stackrel{\star\star}{\beta}
+\frac{3\stackrel{\star}{\beta}^2}{4}
+ \frac{\stackrel{\star\star}{\alpha}}{2} + \frac{
\stackrel{\star}{\alpha}^2}{4} +
\frac{\stackrel{\star}{\beta}\stackrel{\star}{\alpha}}{2}
-
\frac{\stackrel{\star}{\gamma}\stackrel{\star}{\beta}}{2}
- \frac{\stackrel{\star}{\alpha}\stackrel{\star}{\gamma}}{4}\right],\\
G^4_{ \  4} & = & e^{-\alpha} \left[\frac{3\ddot\beta}{2} +
\frac{3\dot\beta^2}{2}-
\frac{3 \dot\alpha \dot\beta}{4}\right] 
+e^{-\gamma}\left[\frac{3\stackrel{\star}{\beta}^2}{4} +
\frac{3\stackrel{\star}{\beta}\stackrel{\star}{\alpha}}{4} \right],
\end{eqnarray}
where overstars and overdots denote respectively ${\partial \over
\partial\psi} $ and ${\partial \over \partial t}$, and
$i=1,2,3$. Following the signature $(-,+,+,+)$ for the
4D metric, we define
$T^0_{ \  0} = -\rho$ and $T^1_{ \  1} = {\rm p}$, where
$\rho$ is the total energy density and ${\rm p}$ is the pressure.
The 5D-vacuum conditions ($G^A_B =0$) are given by\cite{6}
\begin{eqnarray}
&& 8\pi G \rho = \frac{3}{4} e^{-\alpha} \dot\beta^2, \label{6''} \\
&& 8\pi G {\rm p} = e^{-\alpha} \left[\frac{\dot\alpha\dot\beta}{2} -
\ddot\beta - \frac{3\dot\beta^2}{4}\right], \label{7''} \\
&& e^{\alpha} \left[\frac{3\stackrel{\star}{\beta}^2}{4} + \frac{3
\stackrel{\star}{\beta}\stackrel{\star}{\alpha}}{4}\right]=
e^{\gamma} \left[\frac{\ddot\beta}{2} + \frac{3\dot\beta^2}{2} - \frac{
\dot\alpha\dot\beta}{4}\right]. \label{8''}
\end{eqnarray}
Hence, from eqs. (\ref{6''}) and (\ref{7''}) and taking
$\dot\alpha = 0$, we obtain
the equation of state for the induced matter 
\begin{equation}\label{9a}
{\rm p} = - \left(\frac{4}{3} \frac{\ddot\beta}{\dot\beta^2} + 1\right)
\rho.
\end{equation}
Notice that for $\ddot\beta/\dot\beta^2 \le 0$ and $\left|\ddot\beta/\dot\beta^2
\right| \ll 1$ (or zero), this equation describes an inflationary universe.
If $\dot\beta =2H_c$ ($H_c$ is the classical Hubble parameter),
the equality $\ddot\beta/\dot\beta^2 =0$ corresponds with a 4D de Sitter
expansion for the universe [metric (\ref{1})].
Inflationary models like a de Sitter expansion or
whose in which $H_c(t) \sim t^{-1}$ [metric (\ref{2})]
can be studied by means of
above approach\cite{4}. However, 
chaotic inflation cannot be studied in this framework.
The generalization of this formalism to inflationary models
with potentials $V(\varphi) \sim \varphi^n $ is one of the
aims of this letter.

\section{Formalism}

In order to develope a different approach to the reviewed 
in the
last section, we can propose
the following metric to describe the universe
\begin{equation}\label{6}
dS^2 = \psi^2 dN^2 - \psi^2 e^{2N} dr^2 - d\psi^2.
\end{equation}
Here, the parameters ($N$,$r$)
are dimensionless and the fifth coordinate
$\psi $ has spatial unities.
As can be demonstrated, the metric (\ref{6}) describes a
flat 5D manifold in apparent vacuum ($G_{AB}=0$).
In the metric (\ref{6})
the parameter $N$ could be a general function of $t$, $r$ and $\psi$
(and perhaps of additional coordinates $\chi_i$ with $i=5,..,n$ and
$d\chi_i =0$), but
in this letter we are going to study the particular case
where $N$ only depends on the cosmic time $t$: $N=N(t)$.
Using the eqs. (\ref{6''}) and (\ref{7''}),
we can calculate
the vacuum solutions of the metric (\ref{6}).
We obtain the following expressions for the 4D induced
pressure (${\rm p}$) and radiation
energy density ($\rho$)
\begin{eqnarray}
&& 8\pi G {\rm p} = -3 \psi^{-2} \label{a}\\
&& 8\pi G \rho = 3\psi^{-2}.\label{b}
\end{eqnarray}
It implies that all the matter (here described by $\rho$) is
given by $\psi$.
More exactly, as the metric (\ref{6}) with
$N=N(t)$ describes a extended spatially flat FRW metric, the
results (\ref{a}) and (\ref{b}) indicate that $\psi^{-1}(N) = H_c(N)$,
where $H_c(N)$ is the classical Hubble parameter (see next section).
Note that the induced 4D equation of state give
us a vacuum one ${\rm p}=-\rho$.

Before study some inflationary example we can discuss the
properties of the metric (\ref{6}).
We consider the geodesic equations for the metric (\ref{6}) in a
comoving frame $U^r = {\partial r/\partial S}=0$.
The relevant Christoffel symbols are
\begin{equation}
\Gamma^N_{\psi\psi}= 0, \quad \Gamma^N_{\psi N}= 1/\psi, \quad
\Gamma^{\psi}_{NN}= \psi, \quad \Gamma^{\psi}_{N \psi}= 0,
\end{equation}
so that the geodesic dynamics ${dU^C \over dS} = \Gamma^C_{AB} U^A U^B$
is described by the following equations of motion
for the velocities $U^A$
\begin{eqnarray}
&& \frac{dU^{\psi}}{dS} = -\frac{2}{\psi} U^N U^{\psi}, \\
&& \frac{dU^{N}}{dS} = -\psi U^N U^N, \\
&&  \psi^2 U^N U^N - U^{\psi} U^{\psi} =1, \label{geo}
\end{eqnarray}
where the eq. (\ref{geo}) describes the constraint condition
$g_{AB}U^A U^B=1$.
From the general solution $\psi U^N  = {\rm cosh}[S(N)]$,
$U^{\psi}=-{\rm sinh}[S(N)]$, we obtain the equation
that describes the geodesic evolution for $\psi$
\begin{equation}
\frac{d\psi}{dN} = \frac{U^{\psi}}{U^N} = -\psi {\rm tanh}[S(N)].
\end{equation}
If we define ${\rm tanh}[S(N)]
=-1/p(N)$,
we obtain
\begin{equation}\label{21}
\psi(N) = \psi_0 e^{\int dN/p(N)}
\end{equation}
for the velocities
\begin{equation}
U^{\psi} = - \frac{1}{\sqrt{p^2(N)-1}}, \qquad
U^N=\frac{p(N)}{\psi\sqrt{p^2(N)-1}},
\end{equation}
where $\psi_0$ in eq. (\ref{21}) is a constant of integration.
The resulting 5D metric is given by
\begin{equation}\label{4d}
dS^2=dt^2-e^{2\int H_c(t)dt} dR^2-dL^2,
\end{equation}
with $t=\int \psi(N) dN$, $R=r\psi$ and $L=\psi_0$ for $H_c(t)=1/\psi(t)$.
With this representation, we obtain the following velocities $ U^A $:
\begin{equation}
U^T=\frac{2p(t)}{\sqrt{p^2(t)-1}}, \qquad
U^R=\frac{r}{\sqrt{p^2(t)-1}}, \qquad U^L=0.
\end{equation}
The solution $|S|={\rm arctanh}[1/p(t)]$ corresponds to a power-law
expanding universe with time dependent power $p(t)$
for a scale factor $a \sim t^{p(t)}$. Since $H_c(t) = \dot a/a$, the
resulting Hubble parameter is
\begin{equation}
H_c(t)=\dot p {\rm ln}(t/t_0) +p(t)/t,
\end{equation}
where $t_0$ is the initial time.

From the above results we can propose that the universe was born
in a state with $S\simeq 0$ (i.e., in a vacuum state ${\rm p}\simeq-\rho$)
and
evolved through the geodesic $|S|={\rm arctanh}[1/p(t)]$ in a comoving
frame $dr=0$, such that the effective 4D spacetime is a FRW metric
\begin{equation}\label{4d1}
dS^2 = dt^2 - e^{2\int H_c(t) dt} dR^2 - dL^2 \rightarrow
ds^2 = dt^2 - e^{2\int H_c(t) dt} dR^2.
\end{equation}
Note that $L$ depends on the initial
value of $\psi$: $L=\psi_0$. In this framework we can define
the $5D$ lagrangian
\begin{equation}\label{l1}
{\cal L}(\varphi,\varphi_{,A}) = -\sqrt{-^{(5)}g} \left[
\frac{1}{2} g^{AB} \varphi_{,A} \varphi_{,B} + V(\varphi)\right],
\end{equation}
for the scalar field $\varphi(N,r,\psi)$ with the metric (\ref{6}).
Here, $^{(5)}g$ is the determinant of the 5D metric tensor in
(\ref{6}) and $V(\varphi)$
is the potential. On the geodesic $|S|={\rm arctanh}[1/p(t)]$ in the comoving
frame $dr=0$, the effective lagrangian for the metric (\ref{4d1}) is
\begin{equation}
{\cal L}\left(\varphi,\varphi_{,A}\right)
\rightarrow {\cal L}\left(\varphi,\varphi_{,\mu}\right) =
-\sqrt{-^{(4)} g} \left[ \frac{1}{2} g^{\mu\nu} \varphi_{,\mu}
\varphi_{,\nu} +V(\varphi)\right],
\end{equation}
where $^{(4)}g $ is the determinant of the metric tensor in the
$4D$ effective FRW background metric (\ref{4d1})
and $\varphi\equiv \varphi(t,R)$.
In this frame the energy 
density and the pressure, are
\begin{eqnarray}
&& 8 \pi G \rho = 3 H_c^2,\\
&& 8\pi G {\rm p} = -(3H_c^2 + 2 \dot H_c),
\end{eqnarray}
with $H_c(t) =\dot a/a$ for a given
scale factor $a(t) \sim t^{p(t)}$.

\section{An application: Semiclassical chaotic inflation}

The inflationary universe scenario asserts that, at some very early
time, the universe went through a superluminical expansion with a scale
factor growing as $a \sim t^{p(t)}$ (with $p \gg 1$). Inflation is needed
because it solves the horizon, flatness and monopole problems of the very
early universe and also provides a mechanism for the creation of primordial
density fluctuations. For these reasons it is an integral part of
standard cosmological model.

To ilustrate the results of the last section we can develope a semiclassical
treatment\cite{Habib} to a chaotic
inflationary model\cite{chao} with a potential
\begin{equation}\label{pot}
V(\varphi) = \frac{m^2}{2} \varphi^2 + \frac{\lambda^2}{24} \varphi^4,
\end{equation}
where $m$ is the mass of the inflaton field and $\lambda \ll 1$ describes
the self-interaction.
The equation of motion for $\varphi$ and the
Friedmann equation [in an
effective 4D FRW metric (\ref{4d1})], are
\begin{eqnarray}
&& \ddot\varphi + 3 H \dot\varphi -\frac{1}{a^2} \nabla^2 \varphi +
V'(\varphi)=0. \label{fi}\\
&& H^2 = \frac{8\pi}{3M^2_p} \left< \frac{\dot\varphi^2}{2} +
\frac{1}{2a^2} \left(\nabla\varphi\right)^2 + V(\varphi) \right>.
\label{fried}
\end{eqnarray}
We can make a semiclassical treatment\cite{starobinsky} for the scalar field
$\varphi=\phi_c(t) +\phi(\vec R,t)$, where $\phi_c(t) = \left<\varphi\right>$
and the small inflaton fluctuations are zero-mean-valued
$\left<\phi\right>=0$.
If the cosmological constant $\Lambda$ is given by
\begin{equation}
\Lambda = \frac{2m^2}{9} + \frac{4m^4\pi}{\lambda^2 M^2_p} + \frac{
\lambda^2 M^2_p}{18^2 \pi},
\end{equation}
the classical Hubble parameter will be related with the classical
potential through the Einstein equation\cite{BCMS}
\begin{displaymath}
V(\phi_c) = \frac{3 M^2_p}{8\pi} \left[ H^2_c - \frac{M^2_p}{12\pi}
\left(H'_c\right)^2 - \Lambda\right],
\end{displaymath}
where $M_p=G^{-1/2}$ is the Planckian mass
and $H_c={8\pi \over 3 M^2_p} [\dot\phi^2_c/2+V(\phi_c)]$. 
The classical
Hubble parameter for the potential (\ref{pot}) is given by
\begin{equation}\label{hub}
H_c(\phi_c) = \frac{\lambda}{3M_p} \sqrt{\pi} \phi^2_c +
\frac{2 m^2 \pi^{1/2}}{\lambda M_p} + \frac{\lambda M_p}{18 \pi^{1/2}}.
\end{equation}
However, the effective Hubble parameter $H$, is given by the
expression\cite{BCMS}
(there is a little mistake in \cite{BCMS} --- the
correct expression is the following:)
\begin{equation}
H(t) = H_c\left[ 1+ \frac{4\pi}{3 H^2_c} \left< \frac{\dot\phi^2}{2} +
\frac{1}{2a^2} \left(\nabla\phi\right)^2 +
\sum_{n=1} \frac{V^{(n)}(\phi_c)}{n!} \phi^n\right>\right],
\end{equation}
where we denote $H_c\equiv H_c(\phi_c)=\dot a/a$
and $V^{(n)}(\phi_c) \equiv \left.{dV(\varphi)
\over d\varphi}\right|_{\phi_c}$.
If the inflaton fluctuations are small, we can make a first order
expansion on $\phi$ for $V(\varphi)$,
and the following approximation is valid
\begin{equation}
H = H_c\left[ 1 + \frac{4\pi}{3H^2_c} \left< \frac{\dot\phi^2}{2}
+\frac{1}{2a^2} \left(\nabla\phi\right)^2\right>\right].
\end{equation}
The treatment of $H$ in the context of semiclassical inflation is
very problematic because the terms inside the brackets include back-reaction
effects\cite{Nambu}. As was demonstrated by Nambu, back-reaction effects
are different on super Hubble and sub Hubble scales.
On sub Hubble
scales such that effects are important and
the effective curvature is increases, but on super Hubble scales
the consequences of back-reaction are no very important.
For this reason, the standard approximation that appears in the literature
(see, for example \cite{Habib,BCMS,Holten})
consists on making $H=H_c$, because
$(\nabla\phi)^2/a^2$ and $\left<\dot\phi^2\right>$ become
negligible on cosmological scales
at the end of inflation.
For simplicity, in this letter we adopt this approximation.

Since $\dot\phi_c=-{4\pi \over M^2_p} H'_c$,
we can describe
the temporal evolution for the spatially homogeneous component of the
inflaton field 
\begin{equation} \label{phi_c}
\phi_c(t) = \phi_0 \  e^{-\frac{\lambda M_p}{6 \sqrt{\pi}}t},
\end{equation}
where $\phi_0$ is $\phi_c(t_0)$.
If we replace  (\ref{phi_c}) in the expression for the Hubble
parameter (\ref{hub}), we obtain its temporal dependence 
\begin{equation}\label{hubt}
H_c(t) = \frac{\lambda \sqrt{\pi} \phi^2_0
e^{-\frac{\lambda M_p}{3\sqrt{\pi}}}t}{3 M_p}
+ \frac{2 m^2 \sqrt{\pi}}{\lambda M_p} + \frac{\lambda M_p}{18 \sqrt{\pi}},
\end{equation}
such that, for a scale factor that evolves as $a(t) \sim t^{p(t)}$
[i.e., $a(N) \sim e^{N}$ in the representation (\ref{6})],
we obtain the differential equation
\begin{equation}\label{df}
\dot p(t) {\rm ln}(t/t_0) + p(t)/t =H_c(t).
\end{equation}
Here, $H_c(t)$ is given by eq. (\ref{hubt}).
If we replace eq. (\ref{hubt}) in (\ref{df}), we obtain the temporal
evolution for $p(t)$
\begin{equation}
p(t) = \frac{18 C M^2_p \lambda \sqrt{\pi} + t\left(36 m^2 \pi M_p +
\lambda^2 M^3_p\right)-18 \pi^{3/2} \phi^2_0 \lambda
e^{-\frac{\lambda M_p}{3 \sqrt{\pi}}t}}{18 M^2_p \lambda \sqrt{\pi}
{\rm ln}(t/t_0)},
\end{equation}
where $C$ is a dimensionless constant of integration.
Note that the last term in the numerator approaches to zero
before inflation ends.

In the figure 1. we show $|S(t)|$ (dotted line) and $p(t)$ (continuous
line) for
$m=0.8 \  10^{-17} \  M_p$ (i.e., $1.5 \  10^2 \  GeV$),
$\lambda = 10^{-15}$, $C=30$
and $\phi_0 =0.1 \  M_p$. Note that $p(t) \rightarrow 1$
at the end of inflation
(i.e., for $t \simeq 10^{13} \  M^{-1}_p$), but $|S(t)|$ increases
from its initial value $S(t_0)=0$.
The interesting here is that the mass value of the inflaton field
agrees with the expected 
for the Higgs mass: $M_{Higgs}\simeq 150 \  GeV$\cite{...}.\\

\section{Final Comments}

We have developed a cosmological model from non-compact Kaluza-Klein
theory, in which the evolution of the early
universe is described by a geodesic
trayectory $|S(N)| = {\rm arctanh}[1/p(N)]$ in a comoving frame $dr=0$
of a 5D metric
\begin{displaymath}
dS^2 = \psi^2 dN^2 - \psi^2 e^{2N} dr^2 - d\psi^2,
\end{displaymath}
such that, by means of the transformation $t=\int \psi dN$, $ R=r\psi$
and $L=\psi_0 $, the resulting 5D background metric for $\psi=H^{-1}_c$
is described by
\begin{displaymath}
dS^2 = dt^2 - e^{2\int H_c(t)dt} dR^2 - dL^2,
\end{displaymath}
which give us an effective 4D FRW background metric
\begin{displaymath}
ds^2 = dt^2 - e^{2\int H_c(t) dt} dR^2,
\end{displaymath}
on the hypersurface $L=\psi_0$.
In this model, the 4D effective dynamics is governed by the temporal
evolution of the fifth dimension.
Physical properties such as the mean energy density
and pressure of matter are well defined consequences of how the extra
coordinate enters the metric. That is, matter is explained
as the consequence of geometry in five dimensions.

To ilustrate the model we have studied
a chaotic inflationary model with $p(N) >1$ for a massive inflaton
field which is self-interacting. An interesting result 
is that the mass of the inflaton field  here obtained
($m=1.5 \  10^2 \  GeV$),
agrees quite well with the expected value for the Higgs mass\cite{...}.
Of course, the method could be applied to other inflationary
models with potentials
$V(\varphi) \sim \varphi^n$.
Moreover, the formalism also could
be developed for more general cosmological models where
$|S(N)| = {\rm arctanh}[1/p(N)]$
would give us the evolution of the universe from its creation to the present
epoch. For example, a cosmological model in which the universe
evolves from a ``big bounce'' was considered
in\cite{WL}. However, this issue go beyond the scope of this letter.

\vskip .2cm
\centerline{\bf{Acknowledgements}}
\vskip .2cm
MB acknowledges CONICET, AGENCIA 
and Universidad Nacional de Mar del Plata
for financial support.\\

\newpage
\begin{displaymath}
\end{displaymath}
\vskip 8cm
\noindent
{\bf Fig. 1.} Evolution of
$|S(t)|$ (dotted line) and $p(t)$ (continuous
line). \\

\end{document}